# MagCeptor: Encoding Broadcast-Addressable Logic into Magnetic Receptors

Sishen Yuan[1]†, Baijia Liang[1]†, Tangyou Liu[1], Yiqing Huang[1], Haoxuan Wu[1], Shuo Xu[1], Hongliang Ren[1]*

[1]Department of Electronic Engineering, The Chinese University of Hong Kong, Hong Kong SAR

†Contributed equally; *Corresponding author. Email: hren@cuhk.edu.hk

**Abstract:**

Selectively operating multiple components while simplifying actuation and control hardware is a common challenge in multifunctional mechanical and fluidic systems. Here, we introduce MagCeptor, an addressable magnetic-cluster architecture in which a shared external field conveys commands and supplies actuation energy. Fixed stators establish local magnetic-field gradients, while field-induced mover reorientation modulates the resulting forces to select guide-constrained mechanical outputs. Field-dependent position–orientation energy profiles explain target actuation, non-target retention and passive reset after field removal. The evaluated configuration achieves a maximum normalized force output of 385 mN/mm$^3$. By coupling finite, addressable strokes to mechanical transmissions, accumulators and locks, we demonstrate coordinated rotation and multistep manipulation with retained task states. Hierarchical node-and-channel addressing further enables selective model-fluid release from distributed receivers in ex vivo tissue. These results establish a receiver-side architecture that combines shared magnetic inputs with task-specific mechanisms to organize mechanical work, retain operation states and access spatially distributed functions.

## Main

Selectively operating different components while simplifying actuation and control hardware is a common design goal in multifunctional mechanical and fluidic systems[1-4]. Examples include reagent routing and chamber selection in microfluidic chips[1], programmed release of stored doses from implantable multi-reservoir devices[5], localized mechanical stimulation for prosthetic tactile feedback[6], and microelectromechanical optical-path addressing in chip-scale LiDAR[7]. Providing each component with dedicated control connections and drive hardware can complicate the interconnection and packaging of dense arrays[1,2,7]. Fluidic multiplexing and electronic row-column addressing reduce this burden by decreasing the number of independent control channels required[1,2,7], while global dynamic drives and boundary-excited waves exploit differences in structural responses to selectively write internal mechanical states[3,4]. These approaches need to balance control simplification with response performance: fluidic multiplexing involves trade-offs among external control valve count, switching steps, and output refresh rate[2], whereas mechanical-state addressing should account for switching-threshold uncertainty and the sensitivity of some transitions to deviations in the input waveform[3,4]. Reducing control connections therefore requires joint consideration of internal integration and local response reliability. Biological signaling offers inspiration for embedding selectivity at the receiver: in endocrine fibroblast growth factor (FGF) signaling, circulating hormones can reach multiple tissues, while receptor and co-receptor composition and ligand-recognition properties help determine local responses[8,9]. Structural studies of ligand-gated receptors further reveal how ligand binding is coupled to functional activation through conformational changes[10]. Inspired by these mechanisms, we propose a broadcast-addressable architecture that embeds response selection in mechanical receivers: a shared medium distributes the input, and input-dependent local interactions reshape the potential-energy landscape to generate the required force and displacement at selected outputs while maintaining the positions of non-target outputs.

Low-frequency magnetic fields combine the ability to act through encapsulation with direct, direction-dependent mechanical coupling, making them an attractive medium for broadcast actuation of embedded mechanical receivers[11-14]. Unlike pneumatic circuits that rely on external pressure supplies and fluidic interconnections[15], magnetic actuation requires no fluidic connection between the field source and the receiver. For receivers enclosed in electrically insulating, nonferromagnetic materials such as common polymers, magnetic actuation requires neither the illumination path needed for optical actuation[16] nor the acoustic coupling path required for ultrasound transmission[17]. At the receiver, the applied field interacts directly with permanent magnetic moments to produce torque, while spatial magnetic-field gradients can generate translational forces, enabling mechanical output without intermediate electrical conversion or heating[11,12,18]. However, a shared magnetic field acts on multiple magnetic components simultaneously. Channel selectivity can be achieved by shaping the external field distribution or exploiting differences in receiver responses[19-21]. Extrinsic gradient design (EGD) creates spatial selectivity by adjusting external magnetic sources[22,23] (Extended Data Fig. 1a), with the attainable field and gradient distributions depending on the source layout, independently controlled inputs, and workspace[24]. Other approaches produce distinct receiver responses through programmed magnetization configurations[25,26] or exploit differences in dynamic response associated with magnetization direction[19]. Reliable operation requires these responses to remain sufficiently distinguishable despite fabrication variability and changes in operating conditions. In compact devices, interactions among magnets also make the forces in each channel dependent on the positions and orientations of neighboring components, requiring the magnetic configuration, motion constraints, and overall layout to be designed together[20]. The key design problem for compact magnetic clusters is therefore to coordinate target force–displacement output, non-target retention, and configuration-dependent magnetic coupling under constraints on external control inputs and device volume, maintaining channel selectivity throughout the mechanical operation.

Here, we introduce MagCeptor, an addressable magnetic-cluster architecture based on intrinsic gradient design (IGD) (Extended Data Fig. 1b), in which a shared external magnetic field conveys commands and supplies actuation energy. Local magnetic-field gradients established primarily by fixed stators generate translational forces, which are modulated by field-induced mover reorientation to produce selective mechanical output (Fig. 1a–c and Supplementary Movie 1). By generating the driving gradients within the receiver rather than relying on gradients from the external source, the architecture achieves a maximum normalized force output of 385 $mN/mm^3$ in the evaluated configuration (Extended Data Fig. 1 and Supplementary Text 1). The movers can rotate and translate along prescribed guide channels, responding to the combined fields of the external source, stators and neighboring movers (Fig. 1a). Field direction and amplitude encode channel commands, while pulse timing defines field application, hold and reset phases. The stator arrangement and magnetization configuration, together with the motion constraints, embed selection rules in the receiver: reorientation changes the target mover's axial magnetic force from retention to outward actuation, while non-target outputs remain positionally retained (Fig. 1b). This physical decoding enables channel selection within the receiver without independently imposed external gradients at each output. When the outward magnetic force overcomes the opposing load and friction, the target mover travels to its prescribed mechanical stop. Upon field removal, the mover reorients under the internal local magnetic field, restoring an axial force directed toward the initial position and enabling passive reset.

Time-resolved energy calculations reveal energy storage and conversion during the actuation–reset cycle (Fig. 1c). The configuration-dependent magnetic potential energy is separated into mover–external-field, stator–mover and mover–mover contributions, alongside the kinetic energy. In the illustrated $\alpha$-channel cycle, applying a 20-mT field initially raises the mover–external-field energy. This contribution subsequently decreases as the mover reorients and translates outward, while the stator–mover energy increases and the mover–mover contribution changes with the cluster configuration. During the field-held outward stroke, the decrease in total magnetic potential energy accompanies a transient increase in kinetic energy, which subsides when the mover reaches the mechanical stop. Field removal eliminates the mover–external-field term; the subsequent decrease in internal magnetic potential energy supports return, accompanied by a second kinetic-energy transient. Reset brings the magnetic-energy contributions close to their initial values. The internal magnets therefore store and release energy through configuration changes, whereas the external field-generating apparatus supplies the net energy for repeated operation. Over a closed receiver cycle in which the mechanical state and recoverable stored energy return to their initial values, net external work supplies load work and compensates for mechanical dissipation.

Magnetic scalar-potential maps visualize the field-dependent magnetic environment (Fig. 1d), while energy-colored position–orientation paths resolve its connection to selective motion (Fig. 1e). Each path is parameterized by the signed guide position $p$, azimuthal angle $\phi$ and polar angle $\theta$, with orientations selected using the energy-minimization procedure (Materials and Methods). Extended Data Fig. 2a,b defines these coordinates and shows angular maps of the system magnetic potential energy at the path endpoints. The corresponding profiles of each mover's magnetic potential energy reveal the energetic tendency along its guide (Extended Data Fig. 2c–e). Without an applied field, outward displacement increases the mover's energy, favoring its initial retained position. Matching 20-mT fields—along $+x$ for $\alpha$, $-z$ for $\beta$ and $-x$ for $\gamma$—promote target reorientation and make outward displacement energetically favorable. Under the same command, non-target profiles continue to favor their initial positions, even when their magnetic moments reorient. Field removal restores the energetic preference for inward return. The distal output positions are defined by mechanical stops rather than unconstrained internal energy minima. These profiles connect field-dependent reorientation with target translation, non-target retention and passive reset within the same magnetic receiver.

We extend these addressable, resettable strokes into composable system functions through mechanical transmission, state retention and hierarchical addressing (Fig. 1f). To convert finite local displacements into cyclic mechanical output, shared field commands are demultiplexed into channel-specific power strokes, which are rectified and combined through a common transmission to produce coordinated rotation (Fig. 3). To preserve the effects of actuation after the magnetic movers reset, we couple their strokes to mechanical accumulators and locks that retain task displacements and configurations. Subsequent commands can then build on these retained states to organize anchoring, positioning, grasping and cutting into multistep operations (Fig. 4). For spatial reuse, we arrange complete MagCeptor receivers into an array: the spatial field-strength distribution relative to activation thresholds selects the receiver node, while field direction selects its functional channel. This hierarchical addressing is demonstrated through selective model-fluid release from millimeter-scale multi-reservoir receivers in ex vivo tissue (Fig. 5). Thus, these integrations establish a transferable receiver-side architecture in which a common magnetic addressing principle is combined with task-specific mechanisms to organize mechanical work, retain task states and access spatially distributed functions under shared external driving.

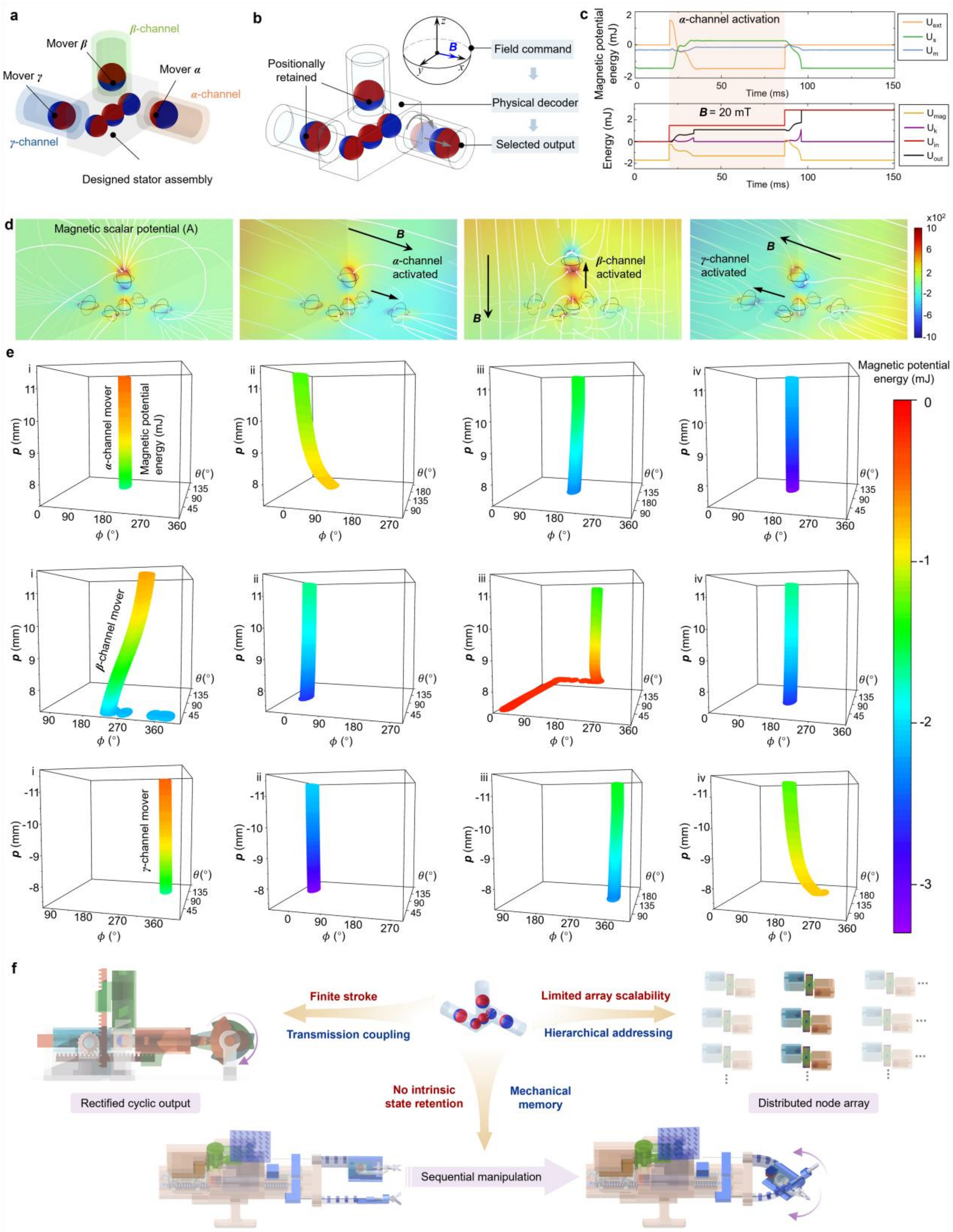

**Fig. 1. MagCeptor architecture, selective magnetic actuation, and system-level functional integration.** (**a**) Three-channel magnetic receiver based on intrinsic gradient design (IGD). A designed assembly of fixed stator magnets establishes local magnetic fields and spatial gradients. The $\alpha$-, $\beta$-, and $\gamma$-movers can rotate and translate along their prescribed guide channels. The translucent channel housings indicate the corresponding mechanical outputs. (**b**) Physical decoding of a shared field command. The direction and amplitude of the external magnetic field, $\boldsymbol{B}$, specify the actuation command. The stator configuration, local magnetic interactions, and motion constraints determine the response of each mover. The illustrated command produces outward translation of the selected mover while the non-target outputs remain positionally retained. Non-target positional retention does not require their magnetic moments to remain fixed. (**c**) Calculated energy evolution during an $\alpha$-channel actuation–reset cycle. The applied field has a magnitude of 20 mT during the shaded interval. The upper plot separates the magnetic potential energy into mover–external-field $U_{ext}$, stator–mover $U_s$, and mover–mover $U_m$ contributions. The lower plot shows the total magnetic potential energy $U_{mag}$, kinetic energy $U_k$, cumulative external energy input $U_{\mathrm{in}}$, and cumulative energy output $U_{\mathrm{out}}$, where $U_{\mathrm{out}}$ is the sum of mechanical work delivered to the load and mechanical dissipation. Field application, mover reorientation and translation, and subsequent field removal redistribute the magnetic interaction energies during outward actuation and return. The external field-generating apparatus supplies energy for repeated operation, while changes in the internal magnetic configuration enable energy storage and release. (**d**) Magnetic-field distributions under four command conditions. Calculated magnetic scalar-potential maps are shown, from left to right, with no applied field and with addressing fields $\boldsymbol{B}_x$ = 20 mT, $\boldsymbol{B}_z$ = –20 mT, and $\boldsymbol{B}_x$ = –20 mT, corresponding to $\alpha$-, $\beta$-, and $\gamma$-channel actuation, respectively. Color denotes magnetic scalar potential in amperes, and white curves indicate magnetic-field lines. Black arrows indicate the applied-field direction and the selected mover's outward motion. (**e**) Energy-minimized position–orientation paths underlying selective actuation. Rows correspond to the $\alpha$-, $\beta$-, and $\gamma$-movers, from top to bottom. Columns i–iv correspond to the four field conditions in (d). Each plot represents the sampled guide position $p$ and magnetic-moment orientation angles $\phi$ and $\theta$; the angular coordinates are obtained using the energy-minimization procedure for the prescribed field and configuration constraints. Color encodes the corresponding magnetic potential energy using a common scale. Without an applied field, outward displacement is energetically unfavorable. Under the matching addressing field, outward motion of the target becomes energetically favorable, whereas non-target outputs retain an energetic preference for their initial positions. The $\gamma$-channel uses a negative signed guide coordinate. Mechanical stops bound the available travel and define the distal output position. The corresponding one-dimensional energy profiles along the guide coordinate are presented in Extended Data Fig. 2. (**f**) Integration of addressable local strokes into system-level functions. Red labels identify receiver-level constraints, and blue labels identify the corresponding integration mechanisms. Left: transmission coupling converts finite, channel-specific strokes into rectified rotary output for power transmission (Fig. 3). Bottom: mechanical memory preserves the effects of earlier commands after the magnetic movers reset, allowing successive operations to build on retained task configurations during sequential manipulation (Fig. 4). Right: hierarchical addressing combines spatial selection of receiver nodes with channel selection within each node, enabling distributed operation through reuse of the local addressing principle (Fig. 5). The schematics illustrate how a common magnetic receiver can be combined with task-specific mechanisms to organize mechanical output, retain operation states, and select operating locations.

**Topology generation and physics-based screening**

We use physics-based topology screening to identify candidate magnetic configurations for selective actuation (Fig. 2a and Supplementary Texts 2–4). Stator-mover-endpoint triplets are combined within a prescribed discrete lattice subject to geometric and occupancy constraints, and duplicate layouts are removed (Supplementary Text 2 and Supplementary Fig. 1). For each layout, the allowed stator magnetization configurations are enumerated. The analytical magneto-mechanical model is then used to calculate the responses of all movers to prescribed field sequences, and candidates are filtered according to their simulated activation patterns (Materials and Methods; Supplementary Text 3 and Supplementary Fig. 2). Retained candidates are compared using activation field, mean non-target holding force, mean repulsive output force and geometric compactness (Supplementary Text 4). Design heuristics from the planar candidate comparisons guide the generation and evaluation of three-dimensional layouts (Supplementary Text 5 and Supplementary Fig. 3).

We implemented a representative three-dimensional MagCeptor with three individually addressable actuation channels (Fig. 2b–d). In this configuration, the $\alpha$- and $\gamma$- channels are aligned antiparallel along the $x$-axis, whereas the $\beta$-channel is aligned along the $z$-axis (Supplementary Text 5 and Supplementary Fig. 3). This opposing bilateral arrangement creates a compact logic gate, with a footprint of three stator diameters, and reduces cross-axis interference by 33–40% relative to the compared configurations (Supplementary Fig. 3). The effects of channel offsets corresponding to 10% of the mover diameter and field-direction deviations of ±20° were assessed through force-response analyses (Supplementary Figs. 4 and 29).

Additional four-, five- and six-channel configurations were evaluated in simulation under the prescribed field sequences (Supplementary Fig. 5). The three-channel configuration was selected for detailed characterization on the basis of its compactness and force output among the evaluated candidates.

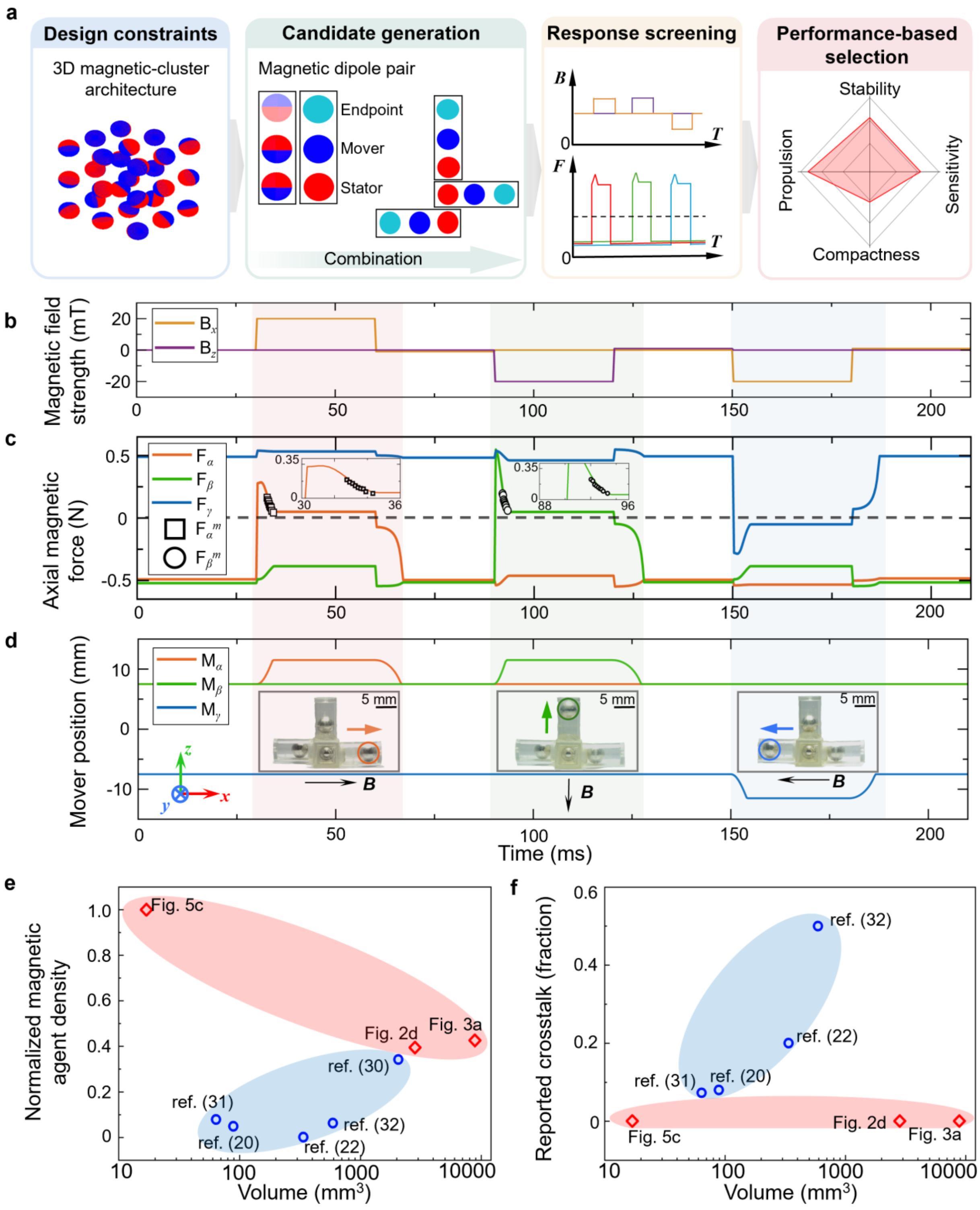

**Fig. 2. Physics-based topology screening and quantitative characterization of MagCeptor. (a)** Candidate-generation and screening workflow. The procedure specifies a discrete design domain and geometric constraints, generates stator–mover–endpoint layouts and enumerates stator magnetization configurations, evaluates the candidate responses to prescribed field sequences using the analytical magneto-mechanical model, and applies the activation–response filter. Retained candidates are compared in activation field, mean non-target holding force, mean

repulsive output force and compactness to select a configuration for further characterization. (**b**) Temporal control sequence. A time-division multiplexed sequence of uniform field pulses (20 mT) is applied along three addressing field directions (+$x$, -$z$, –$x$). (**c**) Calculated magnetic forces and measured axial output forces. Solid curves show the axial magnetic forces calculated for the three movers under the field sequence in (**b**). Open squares ($\mathbf{F}_{\alpha}^{m}$) and open circles ($\mathbf{F}_{\beta}^{m}$) denote experimentally measured $\alpha$- and $\beta$-channel output forces, respectively. The force profiles distinguish target actuation from non-target retention: the addressed mover develops an outward driving force, with predicted peak magnitudes of ~0.28 N for the $\alpha$- and $\gamma$-channels and 0.54 N for the $\beta$-channel, while non-target movers remain retained by holding forces. Force signs follow the indicated spatial coordinate convention. The insets enlarge the comparisons between analytical model predictions and measurements. (**d**) Mover positions and experimental verification of selective motion. Time-lapse photography confirms that global field vectors act as specific keys, triggering only the target mover while non-targets remain locked in potential wells by substantial anchoring forces (>0.3 N). (**e**) Normalized magnetic agent density. The MagCeptor configurations (red diamonds) are compared with selected prior magnetic actuation systems[20,22,30-32] (blue circles) using the defined normalized density metric (Supplementary Table 1). (**f**) Reported crosstalk (fraction). Red diamonds represent MagCeptor configurations (Figs. 2d, 3a and 5c), while blue circles indicate reported or estimated crosstalk from previous studies[20,22,31,32] (Supplementary Table 2). In the unloaded 5,000-cycle reliability test, no non-target false-triggering events were observed (95% Wilson upper endpoint, 0.077%; Supplementary Text 6).

## Characterization

We characterized the selected magnetic configuration under prescribed uniform-field commands (Fig. 2b-d). Under uniform magnetic fields of 20 mT applied along three addressing field directions (Fig. 2b), the MagCeptor exhibits distinct selective responses (exemplified by a capsule-endoscope-sized prototype[27]). For the addressed channel, field-induced mover reorientation produces an outward axial force through local magnetic interactions, with predicted peak magnitudes of ~0.28 N for the $\alpha$- and $\gamma$-channels and ~0.54 N for the $\beta$-channel. Under the same field command, non-target movers remain positionally retained by holding forces exceeding 0.3 N (Fig. 2c,d). The contrasting target and non-target responses demonstrate selective addressing of the three actuation channels.

The dynamic performance of the MagCeptor quantitatively validates predictions from the analytical magneto-mechanical model (Materials and Methods). High-speed tracking recorded mover ejection velocities of 1.7–2.2 m/s, while the analytical model reproduced the measured displacement trajectories with a maximum channel-wise mean absolute error below 0.35 mm (Supplementary Fig. 6a,b). No non-target false-triggering events were observed over 5,000 consecutive actuation cycles with no added external mechanical load (Supplementary Text 6). Analytical modeling under constant opposing loads of 0–0.15 N showed that lower loads allowed the $\alpha$-mover to reach and remain at its travel limit, whereas higher loads reduced the displacement maintained during field application (Supplementary Fig. 6e). Spring-loaded testing included 1,000 actuation–reset cycles for each of the three spring/precompression conditions (Supplementary Fig. 6f). Peak-position variations were below 0.25, 0.30 and 1.0 mm for the 8-, 9- and 10-mm-long springs precompressed by 1, 2 and 3 mm, respectively (Supplementary Text 6).

Systematic geometric scaling studies further validate the robustness of the architecture (Supplementary Text 7 and Supplementary Fig. 6). Increasing the stator and mover dimensions increases both the anchoring force (from 0.68 N to 8.07 N) and the driving force (from 0.62 N to 6.02 N). Notably, as the system scales up, the relative amplitude of force fluctuations on non-target channels is halved from 30% to 15%, indicating a trend where the reduced relative non-target force variation of the MagCeptor improves with scaling. Further scaling analysis confirms that this selective mechanism maintains high fidelity across varying length scales (Supplementary Fig. 6). However, analytical modeling predicts selective actuation and field-off reset in a three-channel receiver with 1-μm-diameter magnetic spheres $R_1$ = 0.5 μm under sequential $\boldsymbol{B}_x$ = +150 mT, $\boldsymbol{B}_z$ = –350 mT and $\boldsymbol{B}_x$ = –150 mT commands ($\mu$ = 0.1; Supplementary Fig. 7c). This prediction is conditional on the prescribed magnetic and contact properties[28,29] (Supplementary Text 8). Independent receiver addressing additionally requires suitable external-field selectivity and target coverage (Supplementary Fig. 24); experimental operation and array density at the predicted scale remain to be established.

The MagCeptor configurations exhibit higher normalized magnetic agent density than the selected prior designs under the defined metric[20,22,30-32] (Fig. 2e, Supplementary Text 9 and Supplementary Table 1). Addressing reliability is evaluated separately through the unloaded 5,000-cycle test (Fig. 2f and Supplementary Text 6). Because prior studies report different selectivity measures, including displacement-based crosstalk and estimated signal overlap, these values are summarized for context rather than interpreted as directly comparable false-triggering probabilities[20,22,31,32] (Supplementary Table 2).

**Wireless physical signal demultiplexing and synchronization**

To demonstrate channel-selective actuation coupled to a common mechanical output, we developed a wireless triple-cylinder engine[33] acting as a physical signal demultiplexer (Fig. 3a,b). Channel selection is implemented through magnetic interactions and motion constraints within the receiver without onboard computation, while the field commands are generated externally. The three MagCeptor channels ($\alpha$, $\beta$, $\gamma$) are mechanically coupled to a common crankshaft through a rack-and-pinion transmission (Fig. 3c; Supplementary Figs. 8–10). Driven by a time-division multiplexed sequence of field pulses (27 mT along $x$, 35 mT along $z$; 10 Hz; Fig. 3d), the MagCeptor architecture functions as a distributed physical decision layer (Fig. 3b). It ensures that only the addressed unit executes a power stroke (latency <25 ms, Supplementary Fig. 11) while non-targeted units remain energetically locked, effectively filtering a serial broadcast into coordinated work (Fig. 3e).

This selective actuation superimposes discrete individual strokes into a coherent macroscopic output (Fig. 3c), generating a robust deterministic synchronization via engineered kinematic redundancy (Supplementary Fig. 10 and Supplementary Movie 2). The transmission generates an output torque of 10.5 N·mm (Fig. 3e and Supplementary Fig. 9), approximately 5.6 times the calculated magnetic torque on a volume-matched single mover under the same applied field conditions[34] (Supplementary Fig. 11). Kinematic characterization demonstrated repeatable stepwise rotation, with median angular increments of approximately 2°–4° across the three channels (Supplementary Fig. 12). This logically gated mechanical transmission enables versatile control, including precise stepwise positioning (Supplementary Movie 3) and programmable bidirectional oscillation (Supplementary Movie 4). No desynchronization was observed over 5 min, and the measured temperature rise reached approximately 4.5°C under the tested conditions (Supplementary Fig. 12 and Supplementary Movie 5). Notably, robustness persists even under nonlinear fields from off-the-shelf permanent magnets (Supplementary Fig. 13, Materials and Methods, and Supplementary Movie 6), validating the architecture's adaptability to unstructured magnetic environments.

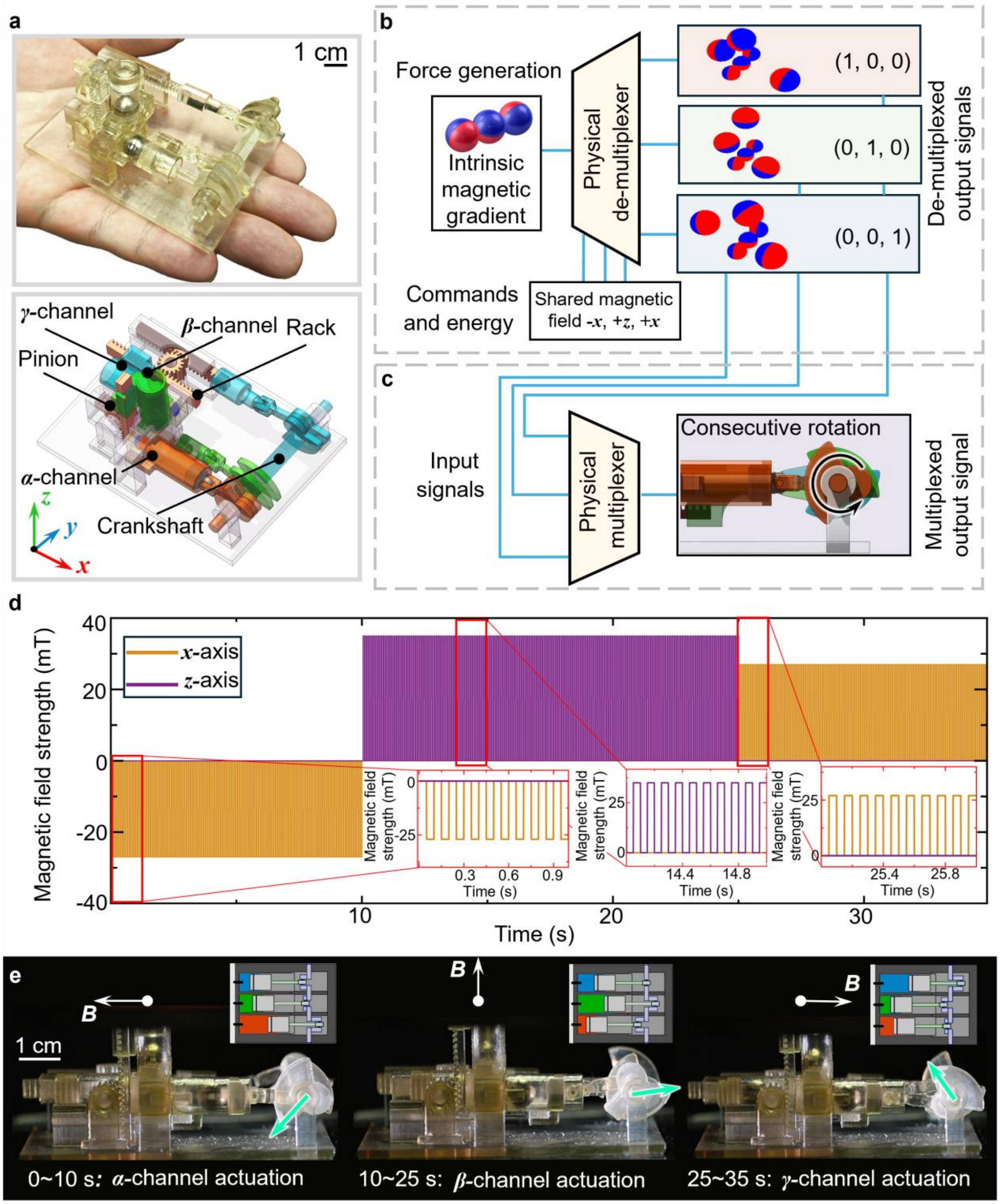

**Fig. 3. Wireless physical signal demultiplexing and synchronization via a triple-cylinder magnetic engine.** (**a**) System architecture. Optical photograph (top) and CAD schematic (bottom) of the prototyped wireless engine. The system integrates three magnetically addressable actuation channels ($\alpha$, $\beta$, $\gamma$) mechanically coupled to a common crankshaft via a rack-and-pinion transmission. Scale bar, 1 cm. (**b**) A shared external magnetic field supplies commands and actuation energy. Local magnetic interactions and motion constraints implement physical demultiplexing, producing channel-selective mechanical outputs. Binary tuples indicate the selected channel in the order ($\alpha$, $\beta$, $\gamma$). (**c**) Macroscopic output reconstruction. Schematic of the physical multiplexer mechanism, where the independent linear strokes of the demultiplexed units are kinematically integrated to rectify discrete magnetic pulses into continuous, synchronized rotary work. (**d**) Control sequence. The time-division multiplexed global broadcast used to drive the engine, displaying magnetic field strength (27 mT along $x$, 35 mT along $z$; 10 Hz) applied along the $x$-axis (orange) and $z$-axis (purple). Insets detail the square-wave pulse structure. (**e**) Operational validation. Time-lapse snapshots verifying the transformation of a serial broadcast into coordinated work. The sequence demonstrates selective addressing phases: the $-x$ field addresses the $\alpha$-channel (0 to 10 s); the $+z$ field addresses the $\beta$-channel (10 to 25 s); and the $+x$ field addresses the $\gamma$-channel (25 to 35 s). White arrows indicate the global field vector (***B***) and the resulting crank rotation. Insets highlight the contrast between the active power stroke and the anchored state of non-target pistons, confirming the high-fidelity execution of the “lock-and-key” logic.

## Sequential logic execution via physical finite state machines

Transitioning from direct kinematics to embodied sequential logic, we instantiated a wireless pipeline robot[35] functioning as a physical finite state machine (FSM) (Fig. 4a). To preserve the result of an addressed stroke during mover reset, we coupled the magnetic receivers to dual-pawl Physical Accumulator modules and mechanical locks (Fig. 4a,b; Supplementary Figs. 14–17). The moving pawl advances the rack during actuation, while the holding pawl retains its displacement during the return stroke (Supplementary Fig. 16). These retained rack positions and the anchoring state provide the physical state variables on which subsequent field commands act. The system is described by the discrete state tuple ($n$, $b$, $m$), where $n$ and $m$ represent the cumulative actuation steps of the $\alpha$- and $\gamma$-channels, respectively, and $b$ denotes the binary buffer state of the $\beta$-channel. Shared magnetic commands thus organize successive operations through retained mechanical states without onboard computation. The dual-pawl accumulator produces 50 mm of cumulative linear displacement against a 0.5-N load through repeated magnetic strokes (Supplementary Movie 7), driving the fabricated variable-stiffness soft arms[36] to bending angles of 205° and 250° (Supplementary Fig. 14c,d). Task displacement therefore accumulates across cycles while each magnetic mover returns within its original guide, separating retained task configuration from actuator reset.

The nonuniform external magnetic field used to generate the towing force ($\boldsymbol{F}_T$) also produces parasitic forces and torques on the robot body. Although manageable during towing, these loads can disturb the robot’s position and orientation during horizontal-channel actuation, requiring mechanical stabilization (Supplementary Fig. 18). The anchoring module (Fig. 4b) is logically gated by the MagCeptor’s $\beta$-channel, which functions as a programmable Physical Buffer gate to actively neutralize mechanical disturbances. This logical interlock yields a substantial anchoring force of ~5 N (Supplementary Fig. 18). Furthermore, the robot’s pipeline-conformal structural design establishes multi-point constraints (Supplementary Fig. 18) that effectively counteract parasitic magnetic torques (up to 22 N·mm). Thus, the $\beta$-channel module physically decouples gradient-driven navigation from vector-encoded manipulation to ensure station-keeping precision (Supplementary Fig. 19 and Supplementary Movie 8). We validated this logic in a multi-stage retrieval mission for objects $\Omega_1$ and $\Omega_2$ (Fig. 4b). The robot executes a prescribed state-dependent sequence by alternating between Free states [e.g., (0, 1, 0)] for navigation and Anchored states [e.g., (5, 0, 0)] for manipulation (Fig. 4c). The successful execution of this sequence confirms the capacity of the MagCeptor to execute state-dependent mechanical operations under spatially confined and perturbed conditions (Supplementary Movie 9).

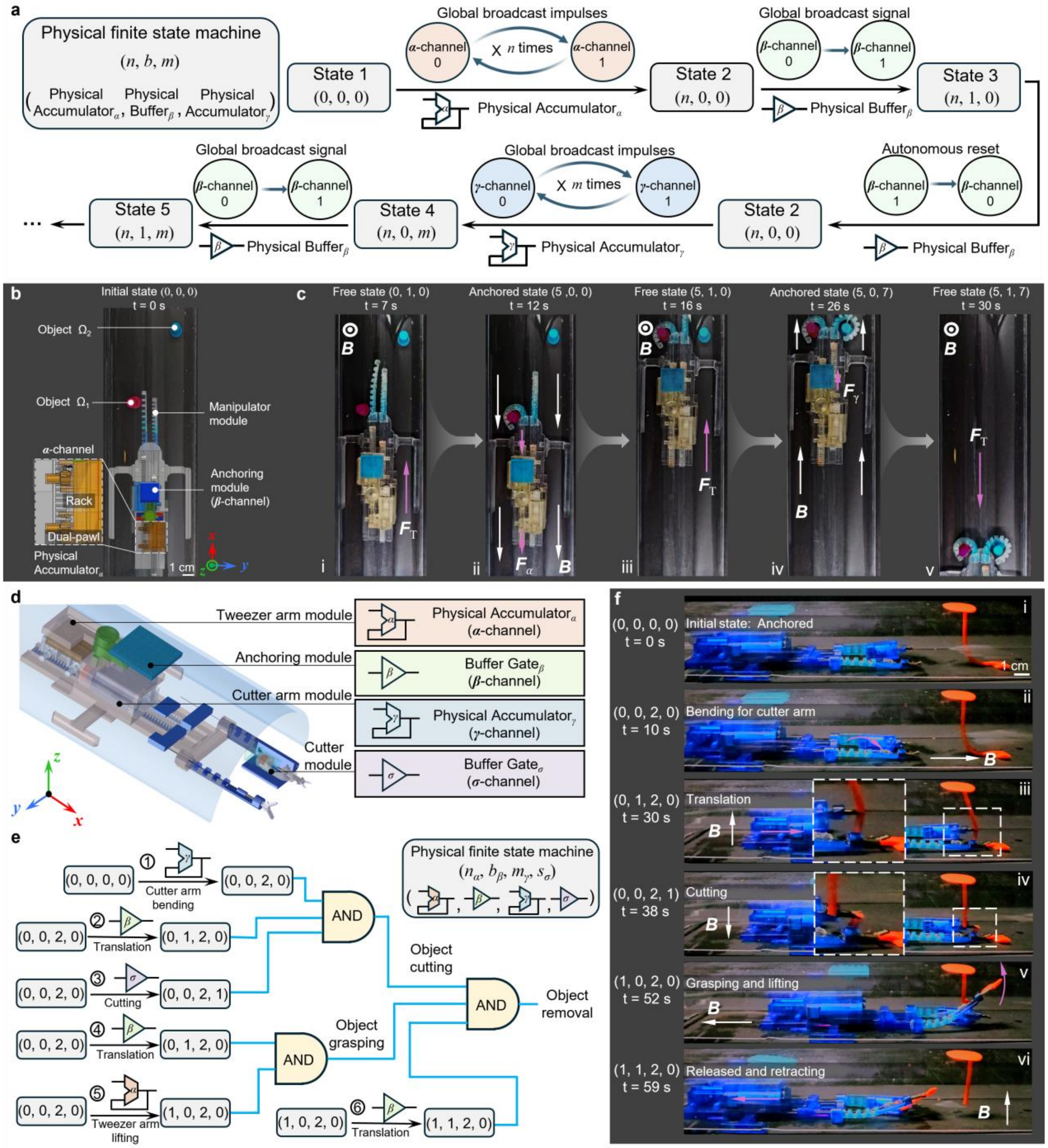

**Fig. 4. Sequential and cooperative logic execution via physical finite state machines.** (**a**) Schematic of the physical finite state machine (FSM) framework. The system state is mathematically defined by the state tuple ($n$, $b$, $m$), representing the accumulated rack steps of the $\alpha$-channel ($n$) and $\gamma$-channel ($m$), and the binary buffer state of the $\beta$-channel ($b$ in $\{0, 1\}$). Arrows denote state transitions deterministically driven by global broadcast impulses. (**b**) Structural instantiation of the wireless pipeline robot incorporating three addressable actuation channels, initialized at state (0, 0, 0). Key functional modules are highlighted, including the manipulator module (actuated by the $\alpha$- and $\gamma$-channels), the anchoring module ($\beta$-channel), and the dual-pawl

physical accumulators. (**c**) Time-lapse sequence of a multi-stage retrieval mission targeting objects $\Omega_1$ and $\Omega_2$. The robot alternates between Free states [e.g., (0, 1, 0)] for towing force ($\boldsymbol{F}_T$) and Anchored states [e.g., (5, 0, 0)] for manipulation. The $\beta$-unit functions as a physical state lock, decoupling the manipulation mechanism ($\boldsymbol{F}_\alpha$, $\boldsymbol{F}_\gamma$) from parasitic loads. (**d**) CAD schematic of the cooperative robot configuration incorporating four addressable actuation channels. The architecture features functional differentiation: the $\alpha$-unit acts as the Tweezer arm (Physical Accumulator$_\alpha$), the $\gamma$-channel as the Cutter arm (Physical Accumulator$_\gamma$), and the $\beta$- and $\sigma$-channels serve as independent Buffer Gate$_\beta$ and Buffer Gate$_\sigma$. (**e**) Task-level logic diagram for cooperative manipulation. The flowchart illustrates the operation of a physical finite-state machine governed by the four-component state tuple ($n_\alpha$, $b_\beta$, $m_\gamma$, $s_\sigma$), corresponding to the states of Physical Accumulator$_\alpha$, Buffer Gate$_\beta$, Physical Accumulator$_\gamma$, and Buffer Gate$_\sigma$. The logical AND operations represent the mechanical-state and task-completion prerequisites for successive operations, including object cutting and retrieval. These conditions organize the externally commanded sequence, while the accumulators and locks provide mechanical state retention. (**f**) Experimental time-lapse sequence of the cooperative task corresponding to the logic circuit in (e). Snapshots (i to vi) capture the robot progressing through discrete tuple states, specifically transitioning from the initial state (0, 0, 0, 0) to the cutting state (0, 0, 2, 1) and finally to the extraction state (1, 1, 2, 0), to sequentially execute anchoring, object resection, and object extraction from t = 0 s to t = 59 s. White arrows indicate the direction of the global magnetic field ($\boldsymbol{B}$), purple arrows denote robot motion, and dashed frames highlight the precise cutting interaction.

### Encoded cooperative logic

To elevate embedded logic from sequential execution to cooperative multitasking, we expanded the MagCeptor architecture to a four-channel configuration (Fig. 4d). This evolution enables functional differentiation among modular subsystems, where the $\gamma$-channel functions as the Cutter arm (Physical Accumulator$_\gamma$) while the $\alpha$-channel acts as the Tweezer arm (Physical Accumulator$_\alpha$). These accumulators integrate a dual-pawl motion rectification mechanism (Supplementary Fig. 16) that functions as mechanical memory, converting volatile magnetic inputs into non-volatile state storage. By mechanically locking the rack against payload tension during field-off intervals, the system retains actuation history without continuous power, thereby satisfying a critical requirement for multistep operation under externally supplied commands in remote environments. The added $\sigma$-channel serves as a decoupled Buffer Gate$_\sigma$ to trigger the distal cutter module. Experimental validation confirms its robust magnetic isolation between limb positioning dynamics and the end-effector triggering mechanism (Supplementary Figs. 20 to 22, and Supplementary Movie 10).

Panels e and f of Fig. 4 present the performance of this cooperative system. Guided by the field of a dual external permanent magnet (D-EPM) source (Supplementary Fig. 13), the robot executes a hierarchical command sequence represented by a physical-state FSM (Fig. 4e). The logical AND operations represent task-level prerequisites rather than mechanically enforced interlocks. As mathematically formalized in the state vector ($n_\alpha$, $b_\beta$, $m_\gamma$, $s_\sigma$), Object cutting requires the prescribed cutter-arm pose ($m_\gamma$) and subsequent activation of Buffer Gateσ ($s_\sigma$), whereas object retrieval requires completion of the cutting operation and engagement of the tweezer arm ($n_\alpha$). These task-level AND relationships organize the externally commanded sequence, with the accumulators and locks retaining the required mechanical states. The time-lapse sequence (Fig. 4f, i–vi) demonstrates coordinated multistep manipulation from initial anchoring to final retraction (Supplementary Movie 11).

Beyond fixed objects, the system demonstrates active cooperative stabilization when addressing unfixed and deformable objects. Here, the logic dictates a synergistic interaction where the stabilizer arm (Physical Accumulator$_{\alpha}$) first engages to grasp and lift the object, thereby establishing an optimal cutting posture to expose the targeted incision point (Supplementary Fig. 21). This alignment facilitates the subsequent shearing action by the Buffer Gate$_{\sigma}$. While these experiments validate kinematic coordination on compliant materials, achieving effective cutting requires finer shear-stress tuning (Supplementary Movie 12), suggesting that future iterations could leverage the scalable discrete topology to achieve more refined actuation dynamics.

**Hierarchical node-and-channel addressing in distributed receivers**

To extend the magnetic addressing principle to millimeter-scale fluidic systems, we developed a multi-reservoir MagCeptor receiver that achieves the highest logic packaging density among the prototypes demonstrated in this work (Fig. 5a). The receiver integrates three separately addressable reservoirs within a 4.5-mm cross-sectional diameter, with the $\alpha$-, $\beta$- and $\gamma$-channels coupled to separate reservoir outputs (Fig. 5b,c; Supplementary Fig. 23). We further arrange complete receivers into a spatial array, with a handheld D-EPM serving as the bus master to supply magnetic commands and actuation energy (Fig. 5d). The spatial field-strength distribution relative to the activation thresholds (~120 mT) selects the receiver node, while field direction selects the reservoir-associated channel within that node (Fig. 5f,g; Supplementary Fig. 24). This hierarchical node-and-channel addressing combines compact within-receiver integration with selective access across distributed operating sites.

High-speed imaging confirms that the high force density yields rapid jet ejection dynamics for a model fluid with a viscosity of 0.03 Pa·s, reaching velocities of ~3.8 m/s within 7.35 ms (Fig. 5c) (Supplementary Fig. 25, Supplementary Text 10 and Supplementary Movie 13). Furthermore, long-duration monitoring in an aqueous environment validates the logic fidelity, exhibiting negligible leakage until the reception of the specific unlock vector key (Fig. 5e). No observable release from non-target reservoirs occurred during the demonstrated addressing sequence (Supplementary Movie 14).

Transitioning from individual nodes to networked topologies, we exploited the nonlinear spatial decay of magnetic fields (Fig. 5f) to establish a three-node array, with three addressable channels per node. By implanting three nodes into tissue at depths of ~5 mm, with ~30 mm spacing, we validated a hierarchical bus addressing protocol (Fig. 5g). This protocol achieves site-selective activation by dual-decoupling the Address Bus (inter-node spatial selection via field magnitude) and the Control Bus (intra-node channel selection via vector orientation). As detailed in the logic truth table (Fig. 5g), aligning the Bus Master over a specific node (e.g., Node 1) with a defined vector (e.g., $+x$) triggers the exclusive release of the corresponding payload (e.g., orange model drug), while neighboring nodes remain energetically locked due to sub-threshold field strengths. Localized release of colored model fluids from selected receivers demonstrates hierarchical node-and-channel addressing in ex vivo tissue (Fig. 5h and Supplementary Movie 15). By enabling externally commanded release without wired connections to the receivers or onboard batteries, the architecture provides an engineering basis for exploring implantable delivery systems and distributed smart implants[37-39].

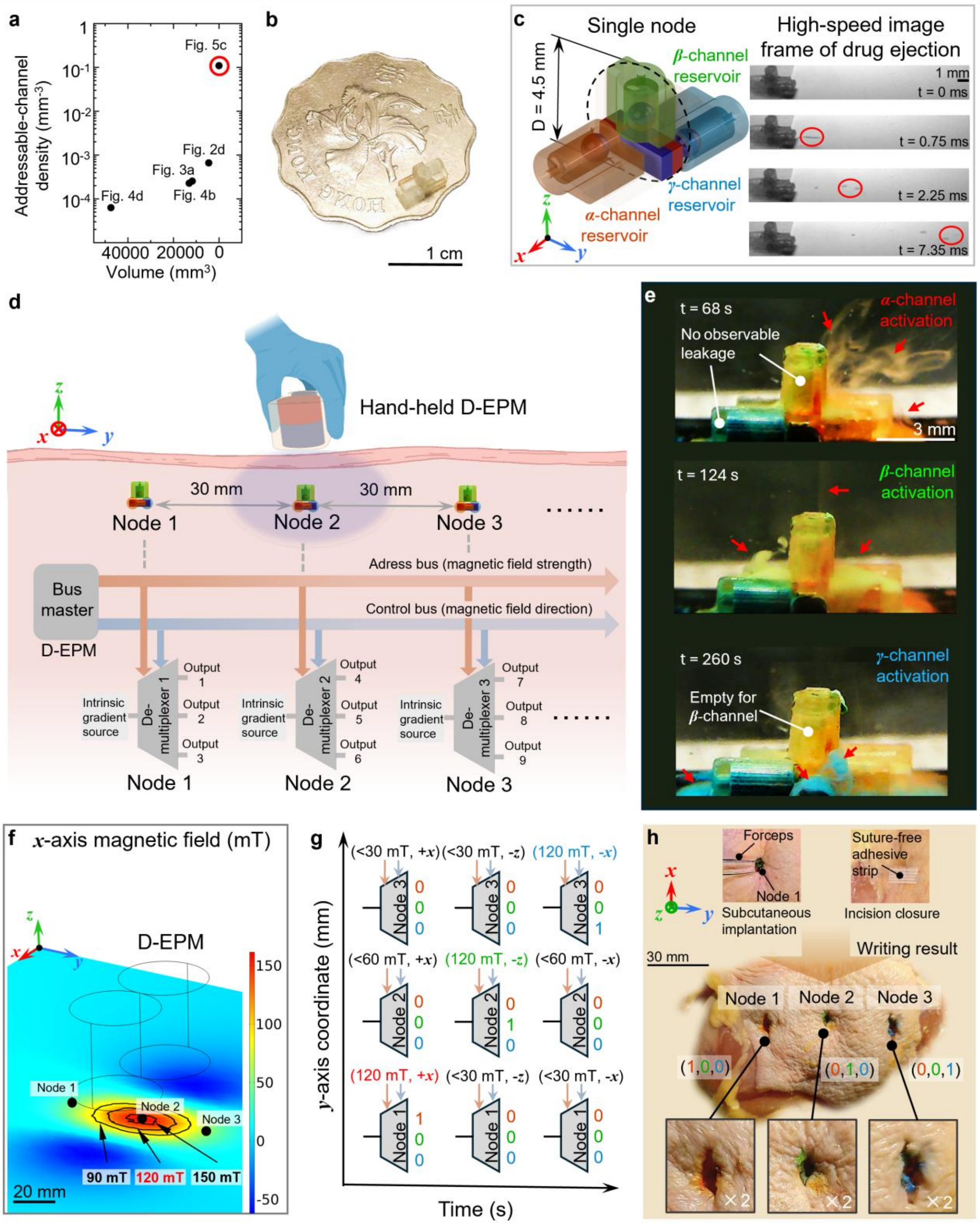

**Fig. 5. Hierarchical addressing and selective model-fluid release in distributed receivers.** (**a**) Benchmarking of logic packaging density. The plot compares addressable-channel density, defined as the number of individually addressable actuation channels per unit volume, across the MagCeptor demonstrators established in this paper (black dots). (**b**) Optical photograph of a fabricated selective drug delivery device alongside a Hong Kong 2-dollar coin for scale. (**c**) CAD schematic (left) of a single node and high-speed dynamic characterization (right). The node

integrates three reservoirs associated with the $\alpha$, $\beta$ and $\gamma$ actuation channels within a cross-sectional diameter of 4.5 mm. High-speed frames capture the ejection of a model fluid with a viscosity of 0.03 Pa·s, reaching approximately 3.8 m/s under the reported actuation conditions. (**d**) Schematic of the hierarchical magnetic addressing protocol instantiated by the D-EPM. Analogous to a digital system, the handheld magnet functions as the Bus Master, utilizing magnetic field strength as the Address Bus (for inter-node spatial selection) and vector orientation as the Control Bus (for intra-node channel selection) to address a distributed network of implants. (**e**) Time-lapse validation of sealing integrity and selective actuation. Sequential $\alpha$-, $\beta$- and $\gamma$-channel release is shown at 68, 124 and 260 s, respectively. Non-selected reservoirs show no observable leakage during the illustrated sequence. (**f**) Finite element simulation of the nonlinear spatial decay of the magnetic field generated by the D-EPM. The distinct gradient between the target node (~120 mT) and neighboring nodes (<90 mT) at 30 mm intervals enables spatial selectivity. (**g**) Logic truth table and timing diagram illustrating the hierarchical addressing protocol. The plot shows how spatially modulating the D-EPM coordinates ($y$-axis), combined with specific vector inputs, triggers exclusive outputs (Logic "1") at target nodes while keeping neighbors in the off state (Logic "0") due to sub-threshold field strengths. (**h**) Ex vivo demonstration of hierarchical node-and-channel addressing. Three receivers, each containing three separately addressable reservoirs, were implanted at a depth of approximately 5 mm with approximately 30-mm spacing. Prescribed D-EPM commands selected the receiver node and reservoir-associated channel, producing localized release of orange, green and blue model fluids.

## Discussion and conclusion

Multicellular organisms achieve sophisticated coordination by embedding information flow directly into a broadcast-addressable molecular architecture. Inspired by this organizational principle, MagCeptor combines command-dependent output selection with axial force generation through stator–mover magnetic interactions, non-target retention and reset. This receiver-side integration contributes to the broader pursuit of physical intelligence through material[40] and structural design[41,42], alongside approaches based on mechanical metamaterials[43,44] and soft pneumatic logic[45]. The IGD architecture uses local magnetic interactions and motion constraints to physically demultiplex shared field commands into channel-specific mechanical outputs. Its addressable strokes couple to transmissions for cyclic motion and to accumulators and locks for retained task configurations, while hierarchical addressing enables localized fluid release from distributed receiver nodes. These combinations connect a common magnetic addressing principle to different mechanical and fluidic functions[46]. Transferability resides in this principle and its integration with task mechanisms, with receiver geometry, field commands, travel limits and load interfaces adapted to each implementation.

The scalability of the MagCeptor architecture is evidenced by its successful translation from coordinated kinematic synchronization at the centimeter scale to millimeter-scale drug delivery. Distinct from passive multistable mechanisms[47], the MagCeptor operates as an active computational primitive capable of both transforming intermittent uniaxial pulses into deterministic quaternary cycles via coupled magnetic selection and morphological constraints (Supplementary Text 11, Supplementary Fig. 26 and Supplementary Movie 16), and realizing physical memory via non-volatile bistable latches (Supplementary Text 12 and Supplementary Fig. 27). The simulated examples include configurations with up to six addressable actuation channels under the prescribed field protocols (Supplementary Fig. 5). The three-node arrangement, with three channels per node, illustrates hierarchical addressing through combined spatial node selection and within-node channel selection (Fig. 5g). For the specified D-EPM, finite element analysis using a nominal 120-mT threshold yields spacing estimates of 13.5 mm along $x$ and 16.5 mm along $y$ (Supplementary Fig. 24a,b). Analytical comparisons show limited inter-node coupling, with normalized mean absolute force deviations below 0.15% in the evaluated two-node arrangements (Supplementary Fig. 24c) and up to 6.37% in a compact 7 × 3.7 mm-pitch array (Supplementary Fig. 24d,e). The central receiver retains channel-selective force responses under virtual target-only excitation. Under the evaluated configurations, these results suggest that external-field selectivity and target-receiver coverage may be more restrictive than inter-node coupling for further array densification. Source miniaturization alone does not satisfy both requirements (Supplementary Fig. 24f).

The mechanical support requirements of MagCeptor depend on how the receiver is integrated into the host system. In fixed installations, the mounting structure can support the receiver and accommodate actuation-related reaction loads. For mobile platforms, the nonuniform driving field can produce parasitic forces and torques that disturb the platform's position and orientation (Supplementary Fig. 18). An addressable channel can operate a reconfigurable anchor to stabilize the platform during manipulation, as demonstrated by the pipe robot (Fig. 4), while bio-inspired adhesive attachments[48] offer a potential alternative. For biomedical applications requiring anatomical localization, mechanical stabilization complements rather than replaces position feedback[49-51]. Ultrasound imaging[52] can provide anatomical referencing, as illustrated by device localization in a porcine esophagus model (Supplementary Fig. 28 and Supplementary Movie 17). Moreover, the MagCeptor exhibits substantial tolerance to control errors, maintaining operational fidelity despite field vector deviations of up to $\pm 20°$ (Supplementary Fig. 29). Notably, this topological stability withstands ferromagnetic interference, preserving logic fidelity despite force modulations up to 33% (Supplementary Text 13 and Supplementary Fig. 30). This provides the necessary control authority to negotiate these field-induced uncertainties.

Looking beyond rigid body mechanics, the MagCeptor architecture offers a foundation for synthesizing fluidic logic circuits. By coupling magnetic selection with fluidic transmission, we demonstrate that force output can be decoupled from the unit's physical orientation (Supplementary Fig. 31 and Supplementary Movie 18). Beyond the single-channel addressing protocols characterized above, selected multi-output responses can be produced using combined field commands (Supplementary Movie 19). Although this coordination involves a trade-off in force efficiency, it grants the multi-state control authority necessary to construct hydraulic and pneumatic logic circuits (Supplementary Fig. 32). This holds potential for implants like programmable hydrocephalus shunts requiring wireless flow regulation[53]. These results support the integration of shared magnetic inputs with task-specific mechanisms for selective mechanical and fluidic operation.

**Acknowledgments:**

The authors gratefully acknowledge Mr. D. Xie, Mr. Q. Liu, Mr. Y. Yang and Mr. M. Su for their dedicated assistance with the experimental setup and data collection. We also thank Mr. T. Zhang, Mr. B. Lin and Dr. J. Lai for their inspiring discussions and insightful suggestions that helped refine the conceptual framework of this work.

**Funding:** This work was supported by the Hong Kong Research Grants Council (RGC) through the Collaborative Research Fund (CRF C4026-21GF), the General Research Fund (GRF 14203323), and the Research Impact Fund (RIF R4020-22); in part by the NSFC/RGC Joint Research Scheme 2022/23 (N_CUHK420/22); and by the Guangdong Basic and Applied Basic Research Foundation (Grant No. 2021B1515120035).

**Author contribution:** All authors contributed to manuscript preparation and to the discussion and interpretation of the data.

Conceptualization: S. Yuan, B. Liang, H. Ren.

Methodology: S. Yuan, B. Liang.

Investigation: S. Yuan, B. Liang, T. Liu, Y. Huang, S. Xu.

Visualization: S. Yuan, B. Liang, T. Liu, H. Wu.

Funding acquisition and administration: H. Ren.

Supervision: H. Ren.

Writing – original draft: S. Yuan and B. Liang.

Writing – review & editing: H. Ren.

**Competing interests:** The authors declare that they have no competing interests.

**Data, code, and materials availability:** All data needed to evaluate the conclusions in the paper are present in the paper and/or the Supplementary Materials.

**Supplementary Materials**

Materials and Methods

Supplementary Text

Supplementary Figs. 1 to 33

Supplementary Tables 1 and 2

References

Supplementary Movies 1 to 19

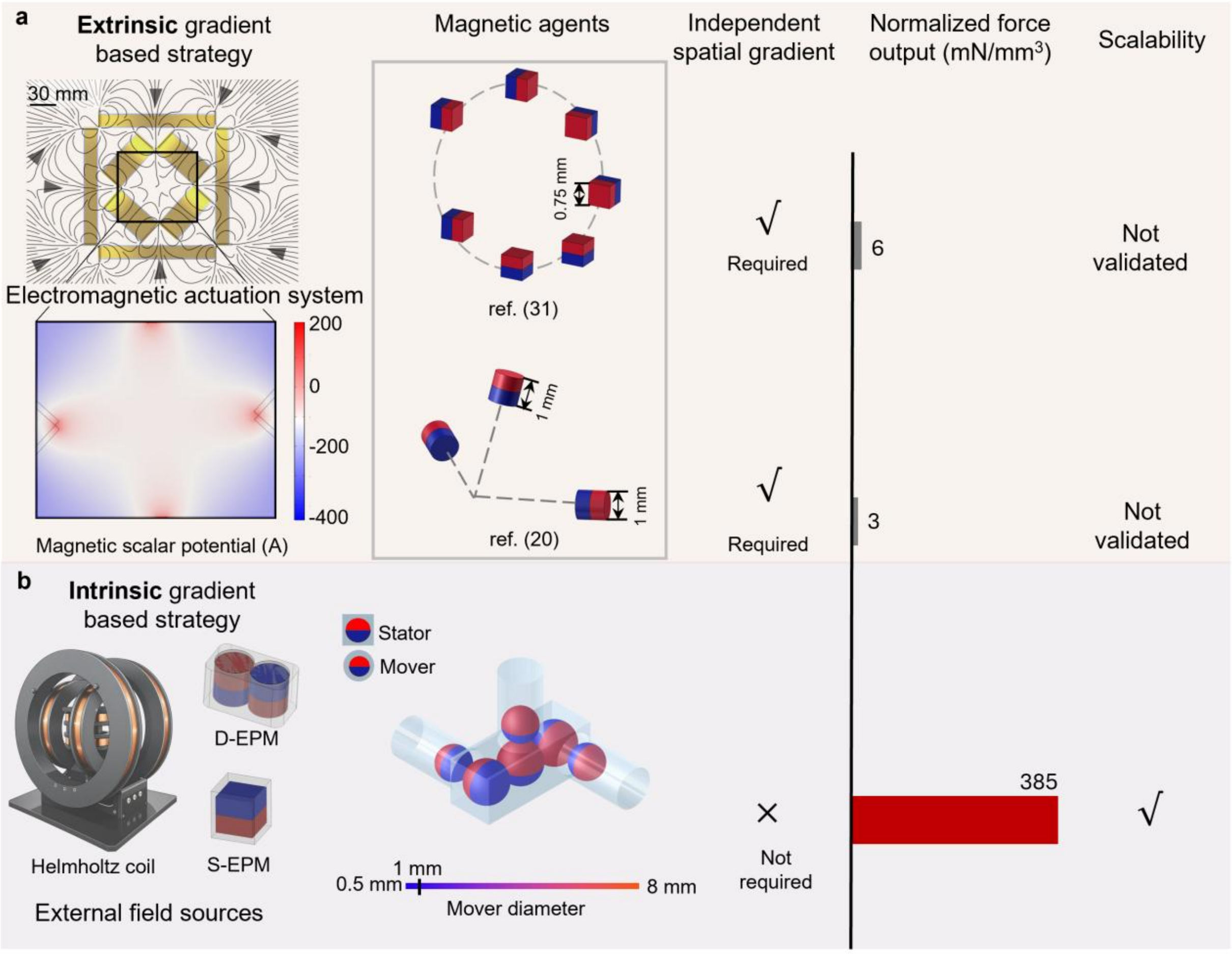

**Extended Data Fig. 1. Comparison of extrinsic and intrinsic gradient strategies for magnetic actuation.** (**a**) Extrinsic gradient design (EGD). External electromagnetic sources generate spatial magnetic-field gradients to actuate distributed magnetic agents. The left panels show a representative electromagnetic actuation system and its magnetic scalar-potential distribution. The middle panels illustrate representative magnetic-agent configurations reported in previous studies[31,20], with characteristic dimensions of 0.75 and 1 mm, respectively. The comparison summarizes the requirement for externally imposed spatial gradients, normalized force outputs of 6 and 3 mN/mm$^3$, and the dimensional-scaling validation status of the selected examples. "Not validated" denotes the status assigned to these examples in the present comparison. (**b**) Intrinsic gradient design (IGD). Fixed permanent-magnet stators integrated into the MagCeptor receiver establish local magnetic fields and gradients. A shared external field reorients the movers, enabling selective translation along their prescribed channels without requiring independently imposed external gradients at each output. The illustrated external field sources include a Helmholtz coil system, a dual external permanent magnet (D-EPM) source and a single external permanent magnet (S-EPM) source. The plotted normalized force output is 385 mN/mm$^3$, and the mover-diameter strip indicates the evaluated dimensional range of 0.5–8 mm. The comparison distinguishes locally generated driving gradients from gradients produced by external sources. Here, scalability encompasses receiver dimensional scaling, expansion of addressable channel count and system-level functional integration. Supporting evidence includes size-dependent characterization and analysis (Supplementary Fig. 6 and Supplementary Text 7), simulated four-,

five- and six-channel configurations (Supplementary Fig. 5), and demonstrations of transmission, mechanical memory and hierarchical node-and-channel addressing (Figs. 3–5).

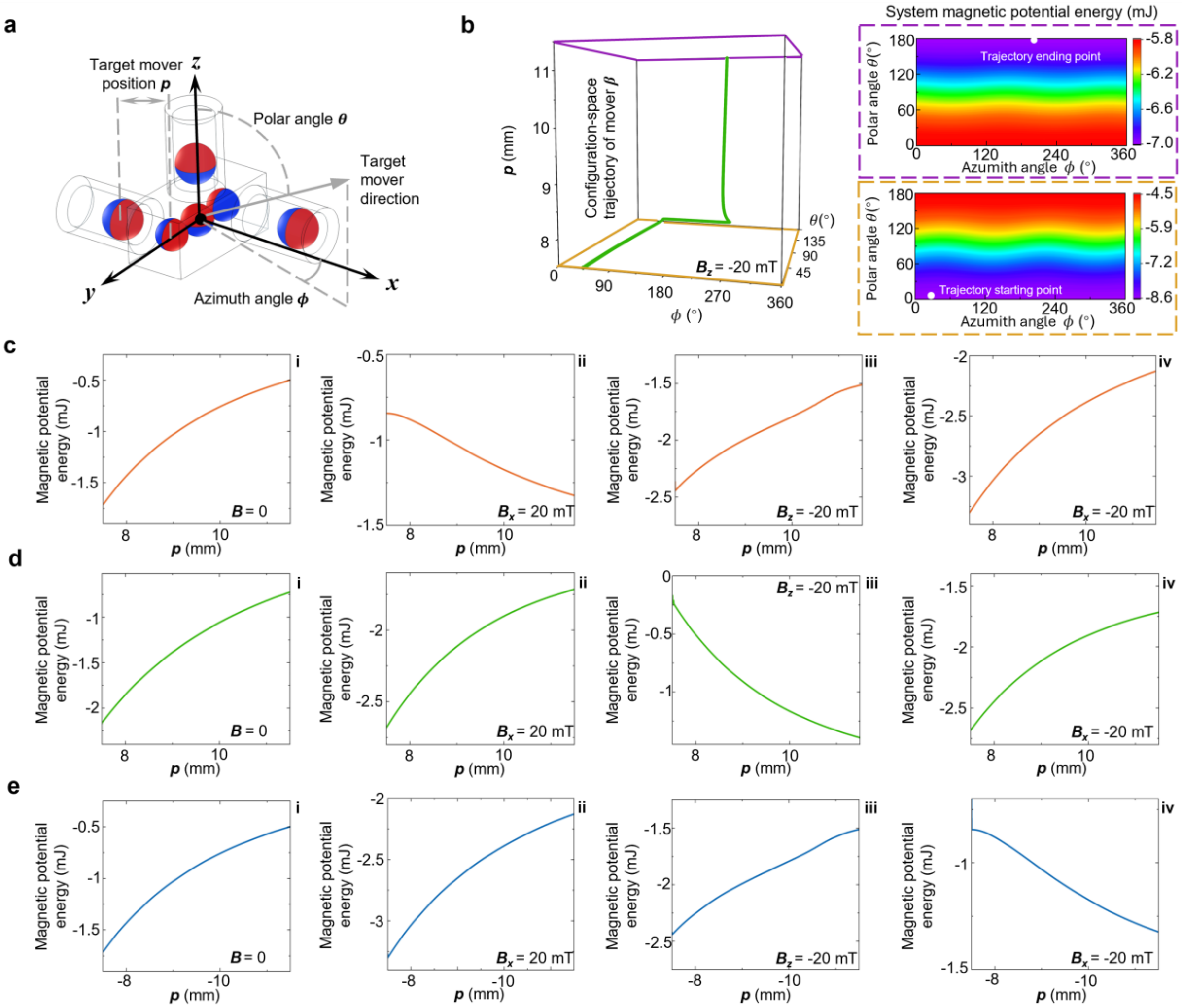

**Extended Data Fig. 2. Position–orientation representation and magnetic potential-energy profiles underlying selective actuation.** (**a**) Definition of the mover configuration coordinates. The target mover's position along its prescribed guide is denoted by $p$. Its magnetic-moment orientation is described by the azimuth angle $\phi$ in the $x$–$y$ plane and the polar angle $\theta$ relative to the $+z$ axis. (**b**) Position–orientation path and endpoint angular-energy maps. The green curve shows the calculated $\beta$-mover path in ($p$, $\phi$, $\theta$) configuration space under $\boldsymbol{B}_z$ = –20 mT. The amber and purple planes indicate the starting and ending guide positions, respectively. The corresponding angular maps of the system magnetic potential energy are shown on the right: the lower, amber-framed map corresponds to the starting position, and the upper, purple-framed map corresponds to the ending position. White markers identify the magnetic-moment orientations at the corresponding path endpoints. Each map uses its own color scale, with energy expressed in millijoules. (**c** to **e**) Magnetic potential energy of individual movers along their guide coordinates. Calculated profiles are shown for the $\alpha$-mover (**c**, orange), $\beta$-mover (**d**, green), and $\gamma$-mover (**e**, blue), corresponding to the position–orientation paths in Fig. 1e. At each sampled guide position, the energy is evaluated at the mover orientations selected by the energy-minimization procedure described in Materials and Methods. Columns i–iv represent the field-off condition and the three addressing fields, respectively: (i) $\boldsymbol{B} = 0$ ; (ii) $\boldsymbol{B}_x$ = 20 mT; (iii) $\boldsymbol{B}_z$ = –20 mT; and (iv) $\boldsymbol{B}_x$ = –20 mT. Without an applied field, the plotted energy increases with outward displacement for all three movers. Under the matching addressing field—(**c**, ii) for $\alpha$, (**d**, iii) for $\beta$, and (**e**, iv) for $\gamma$—the

energy decreases along the outward path, whereas the non-target profiles retain an energetic preference for the initial position. For the $\gamma$-mover, outward travel corresponds to decreasing values of the signed coordinate $p$. The profiles describe the field-dependent energetic preference along the prescribed paths; energy-axis ranges differ between plots.